\documentclass[11pt,a4paper,onecolumn]{scrartcl}
\usepackage[utf8x]{inputenc}
\usepackage[english]{babel}
\usepackage{amsmath}
\usepackage{amsfonts}
\usepackage{amssymb}
\usepackage{graphicx}
\usepackage{threeparttable}
\usepackage[table.xcdraw]{xcolor}
\author{Philipp Koch\footnote{~Correspondence: philipp.koch@ecoaustria.ac.at}~$^,$\footnote{~EcoAustria - Institute for Economic Research, Am Heumarkt 10, 1030 Vienna, Austria.}~$^,\footnote{~Vienna University of Business and Economics, 1020 Vienna, Austria.}~$}
\title{Economic Complexity and Growth}
\subtitle{Can value-added exports better explain the link?}
\usepackage{fancyhdr}
\usepackage{setspace}
\usepackage[olditem,oldenum]{paralist} 
\usepackage{caption}
\usepackage{subcaption}
\usepackage{remreset}
\usepackage{pdfpages}
\usepackage[hidelinks]{hyperref}
\usepackage{longtable}
\usepackage{tabularx}
\usepackage{chngcntr}
\usepackage{threeparttable}

\makeatletter
\@removefromreset{equation}{chapter}
\makeatother

\usepackage{apacite}

\begin{document}

\maketitle

\begin{center}
    \sffamily\bfseries{Abstract}
\end{center}
In economic literature, economic complexity is typically approximated on the basis of an economy's gross export structure. However, in times of ever increasingly integrated global value chains, gross exports may convey an inaccurate image of a country's economic performance since they also incorporate foreign value-added and double-counted exports. Thus, I introduce a new empirical approach approximating economic complexity based on a country's value-added export structure. This approach leads to substantially different complexity rankings compared to established metrics. Moreover, the explanatory power of GDP per capita growth rates for a sample of 40 lower-middle- to high-income countries is considerably higher, even if controlling for typical growth regression covariates. \par
  \par

\section{Introduction}
Economic complexity and its relationship to economic growth is an increasingly researched topic, originating in seminal contributions by \citeA{Hidalgo.2007}, \citeA{Hausmann.2007}, and \citeA{Hidalgo.2009}, who show that economic complexity can explain cross-country income differences and predict future growth rates.
\par
A country is said to be complex if \textit{(i)} it is diversified, i.e. is able to produce and export a wide range of products; and \textit{(ii)} if it exports less ubiquitous products, which are assumed to be more complex than ubiquitous products. By taking both dimensions into account, economic complexity can be seen as a latent measure of the amount of productive knowledge a country holds, which influences economic performance. Hence, complexity is intertwined with, but goes beyond the concept of human capital.  \par 
Existing studies approximate an economy's complexity by its gross export structure in terms of products. However, as \citeA{Koopman.2014} point out, gross exports do not only capture the value that is added in the exporting country, but also the foreign value-added that was imported as intermediate goods, exported goods that are eventually consumed in the domestic country, as well as double-counted exports. Since the share of foreign value added in exports increased over the past decades \cite{Johnson.2017}, gross exports may lead to incorrect conclusions when analyzing a country's economic performance \cite<e.g.>{Timmer.2019}. \par
To take this shortcoming into account, this article contributes to the literature by proposing to approximate an economy's complexity based on the structure of its value-added exports, which are defined as domestic value added in exports of intermediary or final goods that are eventually consumed in a foreign country. The performance of the complexity measure based on value-added exports in explaining GDP per capita growth is then compared to two established indices - the Economic Complexity Index (\textit{ECI}) by \citeA{Hidalgo.2009} and Economic Fitness (\textit{EF}) proposed in \citeA{Tacchella.2012} and \citeA{Tacchella.2013}. 

\section{Data \& Methodology}
\paragraph{Data.} Value-added exports are calculated based on the most recent release of the World-Input-Output Database \cite{Timmer.2015}, covering 56 industries in 43 lower-middle- to high-income annually from 2000 to 2014.\footnote{~A detailed description of the calculation of value-added exports on sectoral level is in \citeA{Koopman.2014}.}$^,$\footnote{~A list of included countries is provided in the appendix. The three countries for which data is not available in every complexity metric, i.e. LUX, MLT and TWN, are excluded from the empirical assessment.} For data on GDP per capita, capital, and population I rely on the Penn World Table \cite{Feenstra.2015}. Data on human capital are obtained from the \citeA{WittgensteinCentreforDemographyandGlobalHumanCapital.2018}. Additionally, data on the $ECI$ are provided by \citeA{TheGrowthLabatHarvardUniversity.2019}, while data on $EF$ can be obtained from the World Bank data catalog.

\paragraph{Methodology.} The $ECI$ and $EF$ are calculated based on a binary adjacency matrix, which denotes whether a country has a Revealed Comparative Advantage \cite{Balassa.1965} in a specific product in terms of gross exports.\footnote{~Specifically, country $c$ has a Revealed Comparative Advantage in product $p$, if $\frac{EXP_{cp}/ \sum_{p} EXP_{cp}}{\sum_{c} EXP_{cp}/ \sum_{c,p} EXP_{cp}} \geqslant 1$.} However, since value-added exports only refer to industries and, thus, the dimensions of the adjacency matrix are reduced considerably, a binary adjacency matrix does not introduce satisfactory variation over time or within cross-sections, since specializations in value-added exports rarely change. Hence, I apply a weighted adjacency matrix \textbf{W}.\footnote{~\citeA{Tacchella.2012} suggest this weighting scheme as a robustness check for the binary adjacency matrix, and refers to the nominator of the Revealed Comparative Advantage \cite{Balassa.1965}.} Its elements are defined as the share of value-added exports ($VX$) country $c$ has in industry $s$, i.e. 
\begin{equation}
\label{eq:W}
    W_{cs} = \frac{VX_{cs}}{\sum_{c} VX_{cs}}
\end{equation}
This adjacency matrix allows for calculating $EF$ in terms of value-added exports, in the following referred to as value-added export Fitness ($VXF$).\footnote{~However, adapting the \textit{ECI} accordingly is not possible, since, by definition of the weighting matrix, $\sum_{c} W_{cs}=1$ $\forall c$. This restricts any variation across industries in the calculation of \textit{ECI} as $k_{s,N}=1$ $\forall N$ \cite<see>{Hidalgo.2009}. Moreover, applying the \textit{ECI} to a binary industry-country adjacency matrix yields results that highly depend on the choice of $N$.} Analogous to \citeA{Tacchella.2012} and \citeA{Tacchella.2013}, it is defined as an iterative process of order $N$ such that
\begin{equation}
\label{eq:1}
    VXF_{c,N}=\frac{\tilde{F}_{c,N}}{\frac{1}{C} \sum_c \tilde{F}_{c,N}}
\end{equation}
\begin{equation}
\label{FitProd}
    Q_{s,N}=\frac{\tilde{Q}_{s,N}}{\frac{1}{S} \sum_s \tilde{Q}_{s,N}}
\end{equation}
$\tilde{F}_{c,N}$ and $\tilde{Q}_{s,N}$ are defined as
\begin{equation}
    \tilde{F}_{c,N}=\sum_s W_{cs}Q_{s,N-1}
\end{equation}
\begin{equation}
\label{eq:4}
  \tilde{Q}_{s,N}=\frac{1}{\sum_c W_{cs}(1/VXF_{c,N-1})}  
\end{equation}
The starting values are set to $\tilde{Q}_{s,0}=1$ $\forall s$ and $\tilde{F}_{c,0}=1$ $\forall c$. $\tilde{F}_{c,N}$ describes the weighted sum of industry complexity levels $Q_{s,N}$, weighted by the share of value-added exports country $c$ has in the respective industry. After normalizing $\tilde{F}_{c,N}$ at every iteration, $VXF_{c,N}$ denotes the complexity level associated with country $c$. The auxiliary variable $\tilde{Q}_{s,N}$ shows that an industry's complexity is positively related to the complexity of countries exporting significant value-added in that industry. Due to the normalization at every step, $VXF_{c,N}$ and $Q_{s,N}$ converge to a unique value for every $c$ or $s$, respectively. \par

\section{Empirical assessment}
In this section, I compare the proposed complexity metric \textit{VXF} to the two established indices based on gross exports. As Table \ref{tab:comparison_rankings} shows for 2014, the complexity country rankings differ substantially. While the United States top the rankings if complexity is approximated by value-added exports, \textit{EF} and \textit{ECI} find China and Japan to be the most complex country, respectively, in 2014. Canada and the Netherlands are, for example, among the top 10 countries in terms of \textit{VXF}, but are not present among the top 10 of the other metrics. For China, the high complexity is driven by a high share of value-added exports in complex manufacturing industries such as electronics, but also by the definition of China in the data, which includes Hong Kong and Macao. This increases the diversity, and thus complexity, considerably due to a high share of value-added in financial services and R\&D. \par

\begin{table}[h!]
\centering
\begin{tabular}{llll}
\textbf{Rank} & \textbf{VXF} & \textbf{EF} & \textbf{ECI} \\ \hline
\multicolumn{1}{l|}{1} & USA & CHN & JPN \\
\multicolumn{1}{l|}{2} & CHN & DEU & DEU \\
\multicolumn{1}{l|}{3} & DEU & JPN & CHE \\
\multicolumn{1}{l|}{4} & GBR & ITA & SGP \\
\multicolumn{1}{l|}{5} & JPN & USA & KOR \\
\multicolumn{1}{l|}{6} & FRA & FRA & AUT \\
\multicolumn{1}{l|}{7} & KOR & ESP & SWE \\
\multicolumn{1}{l|}{8} & ITA & IND & CZE \\
\multicolumn{1}{l|}{9} & CAN & BEL & FIN \\
\multicolumn{1}{l|}{10} & NLD & GBR & HUN
\end{tabular}
\caption{The ten most complex countries for VXF, EF and ECI in 2014}
\label{tab:comparison_rankings}
\end{table}

\begin{figure}[h]
     \centering
     \begin{subfigure}[b]{0.3\textwidth}
         \centering
         \includegraphics[width=\textwidth]{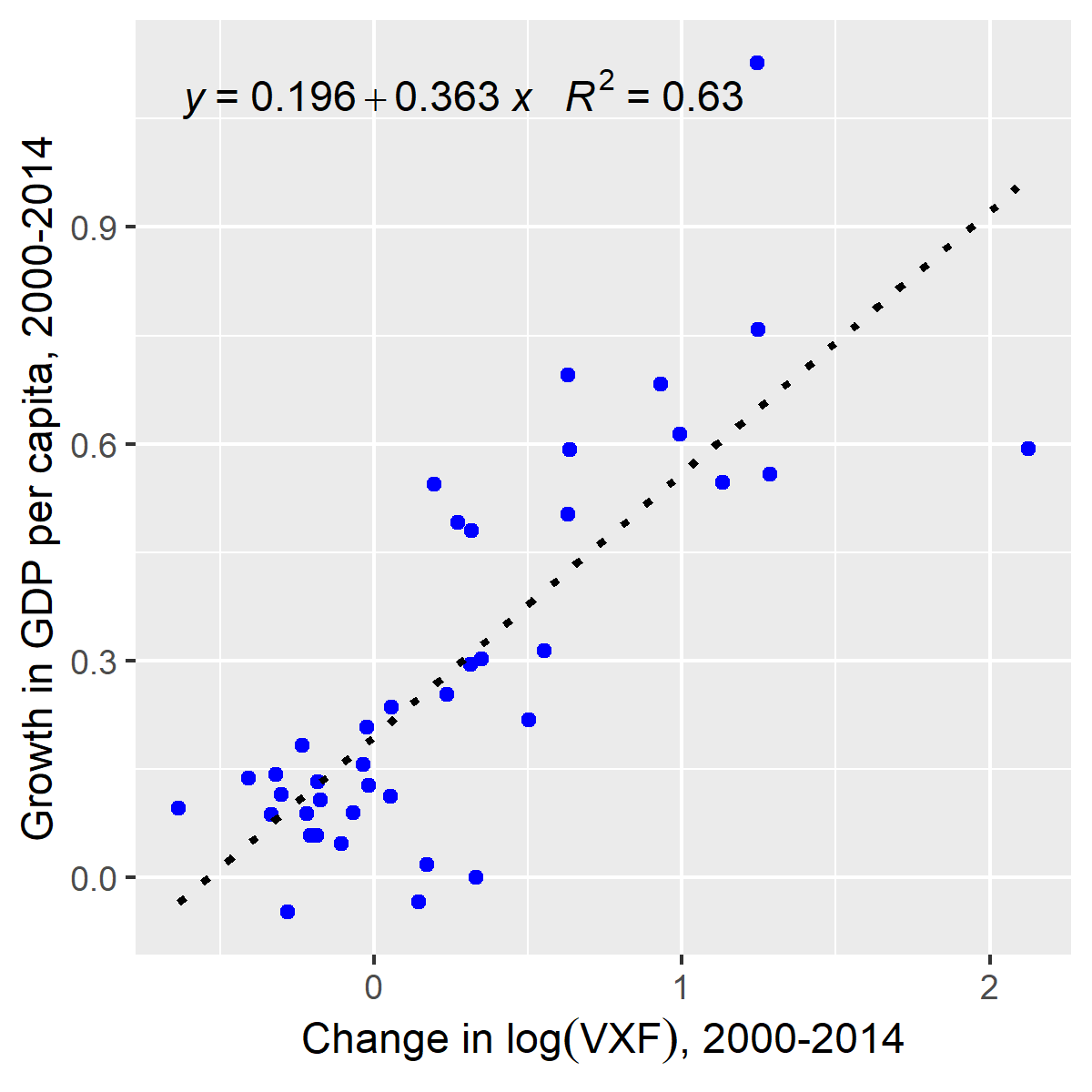}
         \caption{\textit{VXF}}
         \label{fig:uncon_VXF}
     \end{subfigure}
     \hfill
     \begin{subfigure}[b]{0.3\textwidth}
         \centering
         \includegraphics[width=\textwidth]{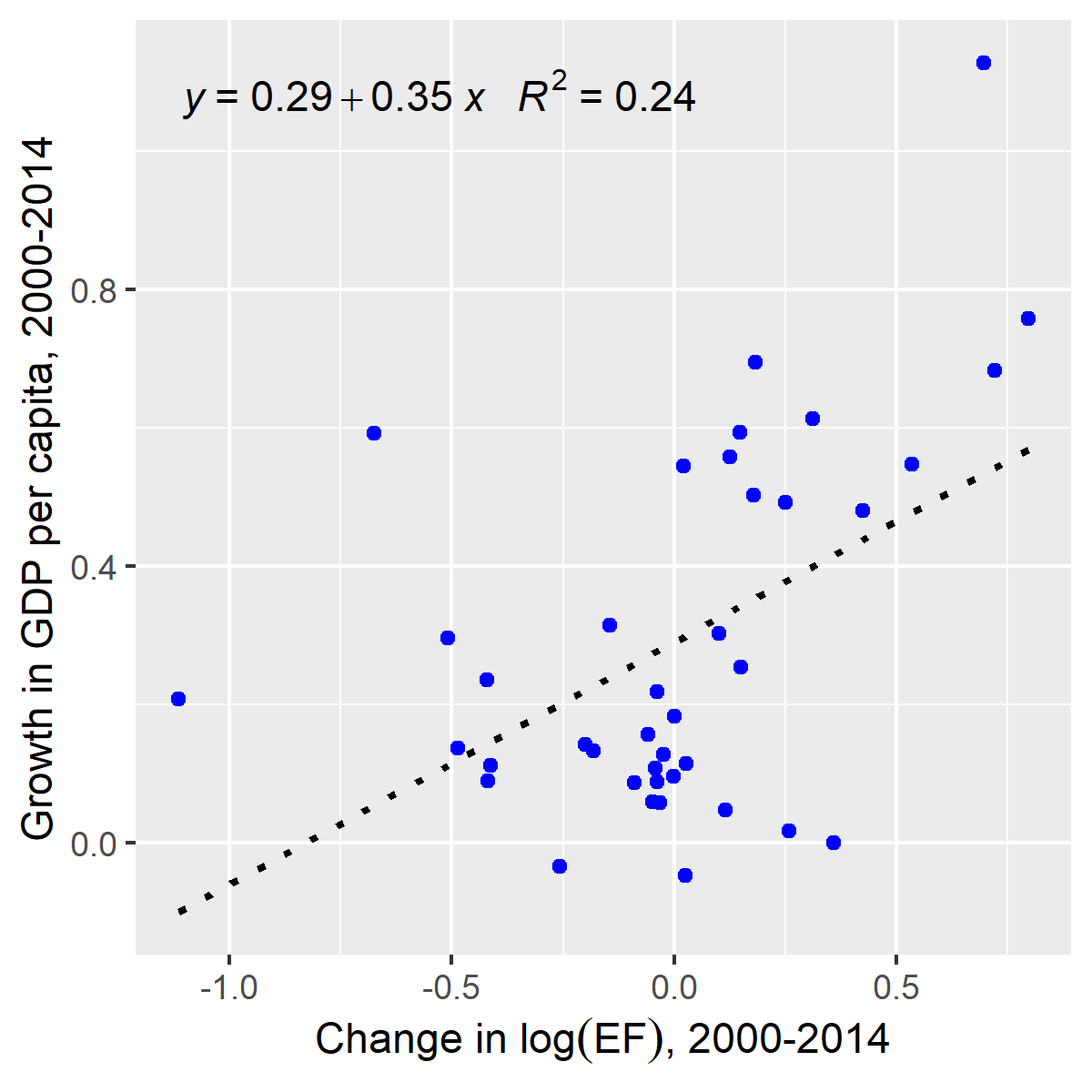}
         \caption{\textit{EF}}
         \label{fig:uncon_EF}
     \end{subfigure}
     \hfill
     \begin{subfigure}[b]{0.3\textwidth}
         \centering
         \includegraphics[width=\textwidth]{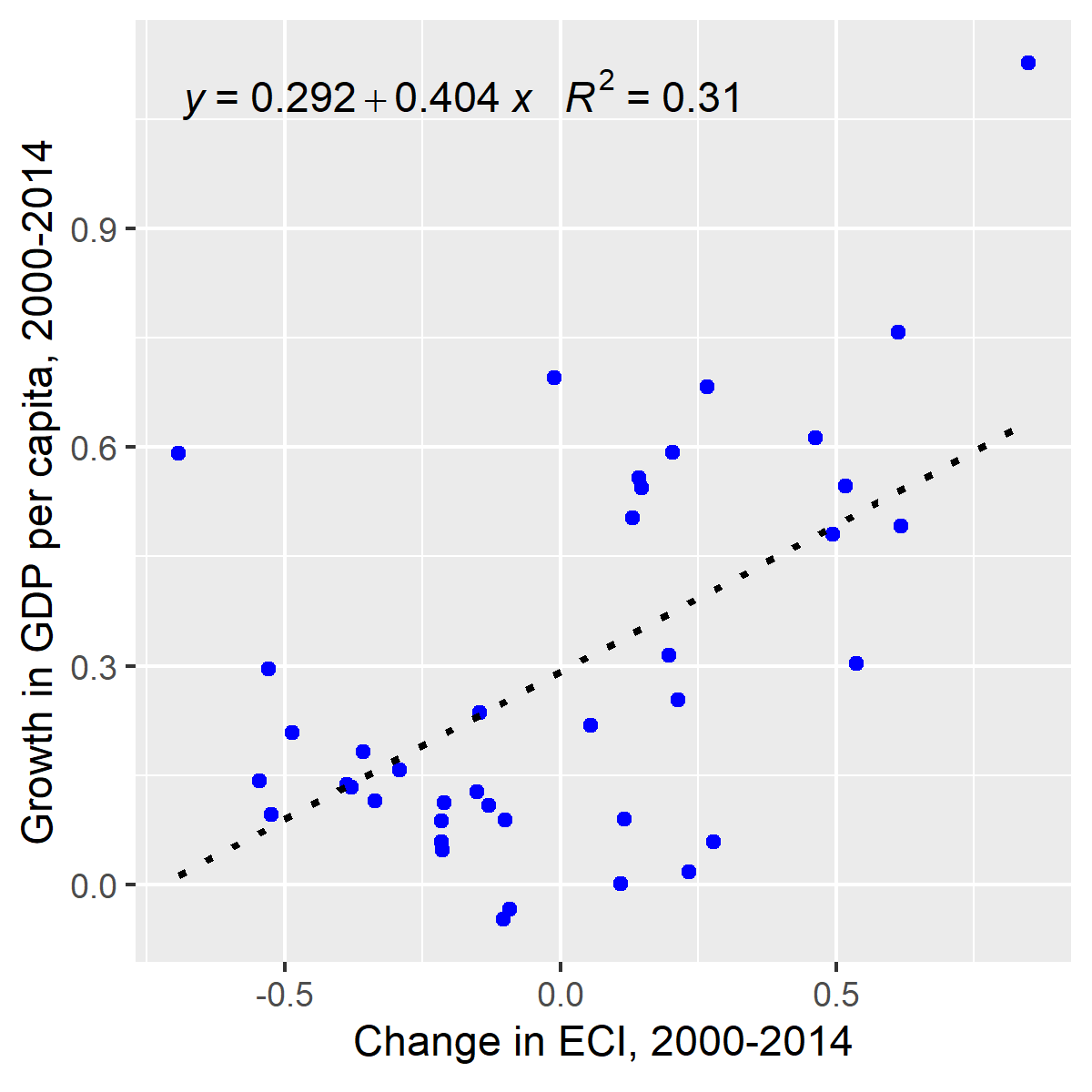}
         \caption{\textit{ECI}}
         \label{fig:uncon_ECI}
     \end{subfigure}
     \caption{Unconditional correlation between complexity and GDP per capita growth}
        \label{fig:uncond.corr}
\end{figure}

The indices seem to vary in their explanatory power of GDP per capita growth rates. For a preliminary inspection, Figure \ref{fig:uncond.corr} displays the unconditional correlation between the growth of each of the complexity metrics and GDP per capita growth between 2000 and 2014. It can be seen that the empirical link between \textit{VXF} and economic growth is stronger with an $R^2$ of $0.63$ compared to $0.31$ ($ECI$) or $0.24$ ($EF$).\par

For further investigation, I create a panel dataset including 40 lower-middle- to high-income countries and comprising three non-overlapping time periods: 2000-2004, 2005-2009, and 2010-2014. I estimate the following first-differenced growth model accounting for individual and time fixed effects
\begin{equation}
     y_{i,t}-y_{i,t-4} = \alpha y_{i,t-4} + \beta_1(C_{i,t}-C_{i,t-4}) + \beta_2 C_{i,t-4} + \gamma X_{i,t} + \theta_t + (\epsilon_{i,t}-\epsilon_{i,t-4}) \quad ,
\end{equation}
where $C$ denotes one of the three complexity metrics and $y$ the logarithmic GDP per capita. $X_{i,t}$ includes typical growth model covariates, i.e. population growth ($n$), change in capital ($K$), change in human capital ($H$), and the initial stock of human capital. \par
Columns (1)-(3) of Table \ref{tab:results} show the regression results. It shows that, on the one hand, the conditional correlation is stronger for $VXF$ than for metrics based on gross exports. On the other hand, including $VXF$ instead of $EF$ or $ECI$ heightens the explanatory power of the model considerably. Moreover, the lower point estimate for the initial human capital stock, if $VXF$ is included, may allow for the conclusion that complexity in terms of a country's value-added export structure more accurately depicts capabilities. \par
Keeping in mind the caveats attached to dynamic panel models, I estimate a non-dynamic fixed effects model as a robustness check.\footnote{~A Generalized Method of Moments (GMM) approach may be preferred, but System-GMM estimations lead to overidentification issues due to the small number of countries.} Columns (4)-(6) in Table \ref{tab:results} show the results. While the impact of complexity in terms of gross exports on growth becomes insignificant in this specification, both for $EF$ and $ECI$, $VXF$ is still significantly correlated with economic growth.\par
Furthermore, the approximation of economic complexity based on value-added exports may be a source of endogeneity, since they directly represent a part of GDP. However, the iterative processes described in Equations \ref{eq:1} to \ref{eq:4} quantify the structure of an economy's value-added exports. That is, an increase in value-added exports does not automatically lead to an increase in $VXF$. Rather, this depends on whether it positively affects the complexity of the value-added export basket, either by an increase in its diversification or by a further specialization in a complex industry (see Equation \ref{eq:W}). To further support that the relationship between $VXF$ and GDP per capita growth is not driven by simply exporting more and thereby increasing value-added, I additionally control for a country's trade openness.\footnote{The applied measure for trade openness is the KOF de-facto trade openness index, which takes exports and imports of goods and services as well as trade partner diversification into account \cite{Gygli.2019}.} Table \ref{tabresults2} in the appendix shows that the results remain qualitatively unchanged, if holding trade openness constant.

\begin{table}[!htbp] \centering 
\footnotesize
\begin{threeparttable}
\begin{tabular}{@{\extracolsep{0pt}}lcccccc} 
\\[-1.8ex]\hline 
\hline \\[-1.8ex] 
 & \multicolumn{6}{c}{\textit{Dependent variable:} $y_{i,t}-y_{i,t-4}$} \\ 
\cline{2-7} 
\\[-1.8ex] & (1) & (2) & (3) & (4) & (5) & (6)\\ 
\hline \\[-1.8ex] 
 $y_{i,t-4}$ & $-$0.254$^{***}$ & $-$0.108$^{***}$ & $-$0.120$^{***}$ &  &  &  \\ 
  & (0.062) & (0.038) & (0.044) &  &  &  \\ 
  & & & & & & \\
  $n_{i,t}$ & $-$0.044 & 0.476 & 0.179 & $-$0.381 & $-$0.193 & $-$0.307 \\ 
  & (0.283) & (0.395) & (0.366) & (0.298) & (0.357) & (0.344) \\ 
  & & & & & & \\
 $\Delta log(K)_{i,t}$ & 0.189$^{***}$ & 0.233$^{***}$ & 0.235$^{***}$ & 0.204$^{***}$ & 0.241$^{***}$ & 0.245$^{***}$ \\ 
  & (0.021) & (0.022) & (0.025) & (0.022) & (0.021) & (0.022) \\ 
  & & & & & & \\
 $\Delta log(VXF)_{i,t}$ & 0.270$^{***}$ &  &  & 0.137$^{***}$ &  &  \\ 
  & (0.044) &  &  & (0.032) &  &  \\ 
  & & & & & & \\
 $log(VXF)_{i,t-4}$ & 0.159$^{***}$ &  &  &  &  &  \\ 
  & (0.030) &  &  &  &  &  \\ 
  & & & & & & \\
 $\Delta log(EF)_{i,t}$ &  & 0.112$^{*}$ &  &  & 0.073 &  \\ 
  &  & (0.068) &  &  & (0.067) &  \\ 
  & & & & & & \\
 $log(EF)_{i,t-4}$ &  & 0.069$^{**}$ &  &  &  &  \\ 
  &  & (0.030) &  &  &  &  \\ 
  & & & & & & \\
 $\Delta ECI_{i,t}$ &  &  & 0.090$^{**}$ &  &  & 0.045 \\ 
  &  &  & (0.036) &  &  & (0.033) \\ 
  & & & & & & \\
  $ECI_{i,t-4}$ &  &  & 0.064$^{**}$ &  &  &  \\ 
  &  &  & (0.031) &  &  &  \\ 
  & & & & & & \\
 $\Delta log(H)_{i,t}$ & $-$0.099 & $-$0.198 & $-$0.045 & $-$0.399 & $-$0.717$^{**}$ & $-$0.644$^{**}$ \\ 
  & (0.304) & (0.339) & (0.312) & (0.293) & (0.301) & (0.287) \\ 
  & & & & & & \\
 $log(H)_{i,t-4}$ & 0.619$^{***}$ & 0.772$^{***}$ & 0.736$^{***}$ &  &  &  \\ 
  & (0.209) & (0.218) & (0.217) &  &  &  \\ 
  & & & & & & \\
\hline \\[-1.8ex] 
Observations & 120 & 120 & 120 & 120 & 120 & 120 \\ 
R$^{2}$ & 0.690 & 0.585 & 0.577 & 0.577 & 0.500 & 0.496 \\ 
Adjusted R$^{2}$ & 0.480 & 0.304 & 0.291 & 0.319 & 0.195 & 0.190 \\ 
\hline 
\hline 
\end{tabular} 
\begin{tablenotes}[flushleft]
    \footnotesize
    \item \textit{Note:} All regressions include individual and time fixed effects. Standard errors accounting for heteroskedasticity are applied. $^{*}$p$<$0.1; $^{**}$p$<$0.05; $^{***}$p$<$0.01
\end{tablenotes}
\end{threeparttable}
\caption{Main regression results} 
  \label{tab:results}
\end{table} 

\section{Conclusion}
This article contributes to the literature by providing a new perspective on complexity. Established metrics approximate an economy's complexity by its gross export structure. I suggest using a country's value-added export structure instead, since value-added exports are arguably a more reliable depiction of economic performance. Based on a weighted adjacency matrix and iterative processes analogous to \citeA{Tacchella.2012} and \citeA{Tacchella.2013}, I introduce the value-added export Fitness ($VXF$) metric. I show that $VXF$, firstly, leads to substantially different complexity rankings compared to the established $ECI$ and $EF$ indices. That is, the United States top the rankings in terms of $VXF$, while Japan and China are the most complex countries in terms of $ECI$ and $EF$, respectively. Secondly, including $VXF$, instead of $ECI$ or $EF$, in a first-differenced growth model with fixed effects controlling for typical growth regression covariates, improves the explanatory power considerably. \par
Further research may rely on different inter-country Input-Output Tables to investigate a larger number of countries, including, specifically, more low- and middle-income countries. A larger sample may allow for a more robust econometric assessment of the effects of complexity on economic growth. Other currently available databases, however, use a coarser industry classification. Moreover, future research should more thoroughly investigate the interactions between human capital and complexity. \par

\section*{Acknowledgements}
I am thankful to two anonymous referees, as well as to Jes\'{u}s Crespo Cuaresma, Wolfgang Schwarzbauer, Johannes Berger, Ludwig Strohner, Martin Wolf, Michael Berlemann, and Tobias Thomas for helpful comments.

\section*{Appendix}
\setcounter{table}{0}
\renewcommand{\thetable}{A.\arabic{table}}

\paragraph{List of countries (ISO 3166-1).} AUS, AUT, BEL, BGR, BRA, CAN, CHE, CHN, CYP, CZE, DEU, DNK, ESP, EST, FIN, FRA, GBR, GRC, HRV, HUN, IDN, IND, IRL, ITA, JPN, KOR, LTU, LVA, MEX, NLD, NOR, POL, PRT, ROU, RUS, SVK, SVN, SWE, TUR, USA.

\begin{table}[!htbp] \centering
\footnotesize
\begin{threeparttable}
\begin{tabular}{@{\extracolsep{0pt}}lcccccc} 
\\[-1.8ex]\hline 
\hline \\[-1.8ex] 
 & \multicolumn{6}{c}{\textit{Dependent variable:} $y_{i,t}-y_{i,t-4}$}  \\
\cline{2-7} 
\\[-1.8ex] & (1) & (2) & (3) & (4) & (5) & (6) \\
\hline \\[-1.8ex] 
 $y_{i,t-4}$ & $-$0.278$^{***}$ & $-$0.150$^{***}$ & $-$0.185$^{***}$ &  &  &   \\
  & (0.057) & (0.037) & (0.037) &  &  &  \\
  & & & & & & \\
  $n_{i,t}$ & $-$0.001 & 0.487 & 0.168 & $-$0.387 & $-$0.265 & $-$0.384  \\
  & (0.271) & (0.377) & (0.356) & (0.290) & (0.354) & (0.357)  \\
  & & & & & & \\
 $\Delta log(K)_{i,t}$ & 0.178$^{***}$ & 0.216$^{***}$ & 0.213$^{***}$ & 0.200$^{***}$ & 0.238$^{***}$ & 0.237$^{***}$  \\
  & (0.021) & (0.021) & (0.022) & (0.022) & (0.020) & (0.021)  \\
  & & & & & & \\
 $\Delta log(VXF)_{i,t}$ & 0.264$^{***}$ &  &  & 0.128$^{***}$ &  &   \\
  & (0.043) &  &  & (0.030) &  &   \\
  & & & & & & \\
 $log(VXF)_{i,t-4}$ & 0.162$^{***}$ &  &  &  &  &   \\
  & (0.028) &  &  &  &  &   \\
  & & & & & & \\
 $\Delta log(EF)_{i,t}$ &  & 0.081 &  &  & 0.044 &   \\
  &  & (0.063) &  &  & (0.062) &   \\
  & & & & & & \\
 $log(EF)_{i,t-4}$ &  & 0.065$^{**}$ &  &  &  &   \\
  &  & (0.026) &  &  &  &   \\
  & & & & & & \\
 $\Delta ECI_{i,t}$ &  &  & 0.118$^{***}$ &  &  & 0.056  \\
  &  &  & (0.037) &  &  & (0.035) \\
  & & & & & & \\
  $ECI_{i,t-4}$ &  &  & 0.087$^{***}$ &  &  &   \\
  &  &  & (0.028) &  &  &   \\
  & & & & & & \\
 $\Delta log(H)_{i,t}$ & $-$0.226 & $-$0.322 & $-$0.242 & $-$0.509$^{*}$ & $-$0.825$^{***}$ & $-$0.740$^{**}$  \\
  & (0.305) & (0.343) & (0.312) & (0.289) & (0.308) & (0.302)   \\
  & & & & & & \\
 $log(H)_{i,t-4}$ & 0.581$^{***}$ & 0.746$^{***}$ & 0.662$^{***}$ &  &  &   \\
  & (0.183) & (0.181) & (0.176) &  &  &   \\
  & & & & & & \\
  $\Delta log(KOF)_{i,t}$ & $-$0.081$^{**}$ & $-$0.135$^{***}$ & $-$0.152$^{***}$ & $-$0.096$^{***}$ & $-$0.116$^{***}$ & $-$0.120$^{***}$  \\
  & (0.035) & (0.035) & (0.036) & (0.036) & (0.035) & (0.036)  \\
  & & & & & & \\
\hline \\[-1.8ex] 
Observations & 117 & 117 & 117 & 117 & 117 & 117  \\
R$^{2}$ & 0.718 & 0.628 & 0.641 & 0.608 & 0.537 & 0.543  \\
Adjusted R$^{2}$ & 0.518 & 0.366 & 0.388 & 0.359 & 0.243 & 0.253  \\
\hline 
\hline 
\end{tabular} 
\begin{tablenotes}[flushleft]
    \footnotesize
    \item \textit{Note} In contrast to Table \ref{tab:results}, these regressions do not include ROU due to availability of the trade openness indicator. All regressions include individual and time fixed effects. Standard errors accounting for heteroskedasticity are applied. $^{*}$p$<$0.1; $^{**}$p$<$0.05; $^{***}$p$<$0.01
\end{tablenotes}
\end{threeparttable}
\caption{Robustness check} 
  \label{tabresults2}
\end{table}

\bibliographystyle{apacite}
\bibliography{MScThesis.bib}

\end{document}